\documentclass[final]{article}

\usepackage[nonatbib]{neurips_2020}

\usepackage[ansinew]{inputenc}
\usepackage{inputenc}
\usepackage{graphicx}
\usepackage{amsfonts}
\usepackage{amssymb}
\usepackage{amsbsy}
\usepackage{amsmath}
\usepackage{multirow}
\usepackage{multicol}
\usepackage{wrapfig}
\usepackage{array}
\usepackage{color}
\usepackage{soul}
\usepackage{cite}
\usepackage[normalem]{ulem} % either use this (simple) or
\usepackage{times}
\usepackage[all]{xy}
\usepackage{lscape}
\usepackage{url}
\usepackage{float}
\usepackage{rotating}
\usepackage{adjustbox}
\usepackage{enumerate}
\usepackage{verbatim}
\usepackage{color,colortbl}
\usepackage{import}
\usepackage{siunitx}
\usepackage{lipsum}
\usepackage[font=small,labelfont=bf]{caption}
\usepackage{float}
\usepackage{caption} 

\usepackage{adjustbox}
\usepackage{changepage}
\usepackage{longtable}

\usepackage[colorlinks=true,citecolor=blue]{hyperref}
\usepackage{booktabs}       % professional-quality tables
\usepackage{amsfonts}       % blackboard math symbols
\usepackage{nicefrac}       % compact symbols for 1/2, etc.
\usepackage{microtype}      % microtypography
\usepackage[colorlinks=true,citecolor=blue]{hyperref}
\usepackage{color}
\newcommand{\x}{{\bf x}}
\newcommand{\y}{{\bf y}}

\newcommand{\remove}[1]{}

\usepackage{graphicx}% Include figure files
\usepackage{dcolumn}% Align table columns on decimal point
\usepackage{bm}% bold math
\usepackage{graphics}
\usepackage{pgf}
\usepackage{tikz}
\usetikzlibrary{arrows,automata,positioning}
\usetikzlibrary{mindmap,trees}

\usetikzlibrary{shapes,arrows,positioning} 

\tikzset{
    >=stealth',
    punkt/.style={
           rectangle,
           rounded corners,
           draw=black, very thick,
           text width=6.5em,
           minimum height=2em,
           text centered},
    pil/.style={
           ->,
           thick,
           shorten <=2pt,
           shorten >=2pt,}
}
\usepackage{array}
\newcolumntype{P}[1]{>{\centering\arraybackslash}p{#1}}

\title{Learning drivers of climate-induced human migrations with Gaussian processes}

\author{Jose M. Tarraga, Maria Piles and Gustau Camps-Valls\thanks{\href{http://isp.uv.es}{http://isp.uv.es}} \\
  Image Processing Laboratory (IPL)\\
  Universitat de Val\`encia\\
  46980 Paterna (Val\`encia). Spain \\
  \texttt{\{jose.tarraga,maria.piles,gustau.camps\}@uv.es}\\
}

\begin{document}

\maketitle

\begin{abstract}
In the current context of climate change, extreme heat waves, droughts and floods are not only impacting the biosphere and atmosphere but the anthroposphere too. Human populations are forcibly displaced, which are now referred to as climate-induced migrants. 
In this work we investigate which climate and structural factors forced major human displacements in the presence of floods and storms during years 2017-2019. We built, curated and harmonized a database of meteorological and remote sensing indicators along with structural factors of 27 developing countries world-wide. We show how we can use \textit{Gaussian Processes} to learn what variables can explain the impact of floods and storms in a context of forced displacements and to develop models that reproduce migration flows. Our results at regional, global and disaster-specific scales show the importance of structural factors in the determination of the magnitude of displacements. The study may have both societal, political and economical implications.
\end{abstract}

\begin{flushright}
\textit{``When we escaped, we ran, without taking our belongings. Some of us ran barefoot. Some women lost their children -- they have seen a lot.''} - Anonymous migrant in Nigeria.
\end{flushright}

\section{Introduction}
An average of 25.3 million displacements has been brought on each year since 2008 due to extreme climate events with devastating consequences on human communities\cite{IDMCreport}. Particularly, vulnerable regions are the most affected with 95\% of climate migrants belonging to low-middle income regions of the globe\cite{W-clim}. Recent studies have shown how changes in climate patterns and the increase in frequency and intensity of climate disasters are inducing newly emerging and changing patterns on human mobility\cite{impact, ipcc}, making migration dynamics more difficult to anticipate.  
In the light of these events, not only novel predictive modelling approaches are needed, but also determining the drivers of climate-induced displacement is crucial to link and quantify climate change impact on migration and to eventually aid effective climate policies and pressure legislative reform\cite{policy}.

The relationship between climate change and human migration is challenging to quantify and disentangle since it is a multi-causal phenomena with complex a interplay between its drivers\cite{clim_era}. Forced displacement occurs when a severity threshold is reached in the affected region\cite{inflexion}, depending on the intensity of the shock but also on a range of socio-economic, demographic, environmental and political factors, among others\cite{factors, effect}. Contrarily to popular narratives, some empirical studies show that greater shock severity does not necessarily need to be proportional to the displacement magnitude\cite{impact}, and displacement is usually modelled by using gravity or radiation equations \cite{gravity, radiation}, which assume linear or log-linear relationships between displacement and the potentially explanatory covariates, calling the need for multivariate and non-linear modelling tools.

While international migration has been more extensively studied, internal displacement has received little attention, probably due to data acquisition challenges. We advocate recent advances in displacement monitoring, together with the increase of airbone, satellite, and meteorological data in sparse-data regions can be exploited for closing this climate-migration gap. Here we propose Gaussian Processes as a machine learning approach to learn from available data how to model internal displacements induced by flood and storm events.

\section{Gaussian processes for modeling and interpretability}
Gaussian Processes (GPs) are non-parametric probabilistic approaches for machine learning problems, mainly for regression and classification. The goal of the GP regression  method~\cite{rasmussen06} is to learn a nonparametric function $f$ able to estimate our target variable (internally displaced person, IDP\footnote{IDP is defined as someone who is forced to flee his/her home but remains within his/her country's borders.}) at country level $y\in\mathbb{R}$ from a set of $D$ input features (e.g. satellite,  meteorological and structural variables), $\x=[x^1,\ldots,x^D]\in\mathbb{R}^{D}$. We assume an additive noise model $y = f(\x) + \varepsilon$, where the noise is additive independent and identically Gaussian distributed with zero mean and variance $\sigma_n$, $\varepsilon\sim{\mathcal N}(0,\sigma_n^2)$. Let us define the stacked output values $\y = [y_1,\ldots,y_N]^\top$, and denote the test points and predictions with a subscript asterisk $\x_\ast$ and $y_\ast$ respectively. The output values are distributed as a a Gaussian with a zero mean and a covariance matrix ${\bf K}$ of size $N\times N$ that contains all pairwise similarities between countries $i$ and $j$, $[{\bf K}]_{ij} = k(\x_i,\x_j)$. 
The GP model prediction for a test point ${\bf x}_\ast$ is obtained by computing the posterior distribution over the unknown output $\y_\ast$ as $p(\y_\ast|{\bf x}_\ast, {\mathcal D})$, where ${\mathcal D}\equiv\{{\bf x}_n,y_n\}_{n=1}^N$ is the training dataset. This posterior can be shown to be a Gaussian distribution, $p(y_\ast|{\bf x}_\ast,{\mathcal D})$ = ${\mathcal N}(y_\ast|\mu_{\text{GP}*},\sigma_{\text{GP}*}^2)$, for which one can estimate the {\em predictive mean} (point-wise predictions) as $\mu_{\text{GP}*} = {\bf k}_\ast^\top({\bf K} + \sigma_n^2{\bf I})^{-1}{\bf y} = {\bf k}_\ast^\top\boldsymbol{\alpha}$, where $\boldsymbol{\alpha}$ are model weights. 

GPs are not black boxes, they allow not only modeling but also gaining some insights about the problem. Here we use a composite covariance formed by a linear and the automatic relevance determination (ARD) kernel function, $K(\x_i,\x_j) = \x_i^\top\x_j + \nu \exp(-\sum_{d=1}^D \gamma_d(x_i^d-x_j^d)^2) + \sigma_n^2 \delta_{ij},$
where $\nu$ is a scaling factor,$\gamma_d$ are dedicated hyperparameters controlling the spread of the signal relations in each dimension $d$, $\sigma_n$ is the noise standard deviation, and $\delta_{ij}$ is the Kronecker's symbol. The linear kernel copes with linear features and to mimic the best linear decision, the 
(anisotropic) exponential kernel deals with locality and nonlinearities to modify the linear solution, and the noise term to regularize the solution. 
The $D+2$ hyperparameters of our GP model $\boldsymbol{\theta} = \{\nu, \gamma_d, \sigma_n\}$ were inferred by Type-II Maximum Likelihood. 
We work in a very data-limited high-noise regime which induces an error surface with many local minima. Therefore, instead of using the conjugate gradient ascend strategies, we used a more robust Bayesian optimization procedure~\cite{Gelbart14}. After optimization, we studied the $\gamma_d$ for model interpretability. 

\section{Data collection and preprocessing}
\begin{wrapfigure}{r}{7cm}
\vspace{-0.25cm}
\centering
  \includegraphics[width=6cm]{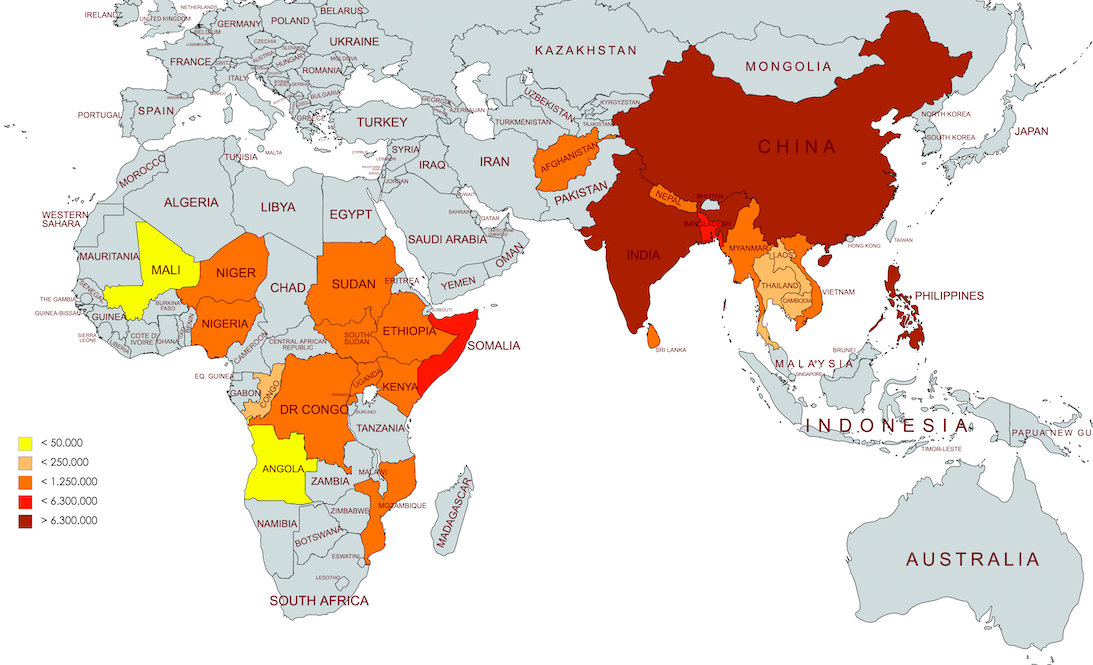}
  \caption{Total IDPs (2017-2019) by country\label{fig:map}.}
  \vspace{-14pt}  
\end{wrapfigure}

The database contains displacement and climatic data from severe floods and storms events in $27$ countries during the period $2017-2019$ alongside a wide range of structural factors in each country. See Fig.~\ref{fig:map} for the total number of IDPs per considered country. 
Displacement data comes from the \href{https://www.internal-displacement.org/}{Internal Displacement Monitoring Centre} (IDMC) which provides Internally Displaced Person (IDP) (our target variable) data on a disaster event basis since $2017$\cite{IDMCweb}.  
We collected a total of $229$ IDP events belonging to Africa and Asia regions and identified the affected areas within each country for each catastrophe consulting different disaster sources, such as IDMC, Relief and EM-DAT \cite{IDMCweb, relief, emdat}. According to the IDMC, $149$ of these events correspond to flood disasters (seasonal rains or displacements of large masses of water) and $80$ of them to storm disasters (tropical cyclones, torrential rain etc). The number of events in each country and the magnitude of displacement are summarized in Appendix I. 

To characterize the {\em severity of the disaster} we extracted climate extremes using the Google Earth Engine platform. The chosen products provide data in sparse data regions, where meteo-hydrological data is poorly available such as Sub-Saharan Africa, from the simulation and combination of satellite and {\em in situ} data \cite{FDLAS1, era5, terra}. 
We selected the maximum valued pixel for each climate feature in the affected area with a monthly resolution. The dates of IDP events and climate extremes were matched by consulting the peak of the catastrophe in different disaster sources \cite{IDMCweb, relief, emdat}. 
Lastly, $46$ variables used in this study  
comprising socio-economic, demographic, political and environmental features were collected from different sources as shown in Appendix II.  
Before training, we standardized data so that $\gamma_d$ was not inflated towards high-variance variables, 
and a log-scale was applied to IDPs and to the total population in the affected area ({\sf \small Pop}).

\section{Results}

\begin{wraptable}{r}{9cm}
%\begin{table}[]
\small
 \renewcommand{\arraystretch}{0.8}
  \caption{Mean and standard deviation results of the regression problems: Pearson's correlation coefficient, r$^2$, mean error, ME, and root-mean-square error, RMSE over 100 runs of 75-25\% train-test partitions. }
  \label{tab:results}
  \centering
  \begin{tabular}{lcccc}
    \toprule
    Models & $N$     & {r$^2$} & {ME} & {RMSE} \\
    \midrule
    Africa & $93$ & $0.56\pm 0.11$ & $0.15\pm 0.09$ & $0.84\pm 0.10$   \\
    Asia & $136$ & $0.64\pm 0.09$ & $0.12\pm 0.09$ & $0.78\pm 0.09$   \\
    \hline
    Flood & $149$ & $0.68\pm 0.08$ & $0.12\pm 0.08$ & $0.73\pm 0.08$  \\
    Storm & $80$ & $0.65\pm 0.13$ & $0.16\pm 0.10$ & $0.76\pm 0.12$   \\
    \hline
    Global & $229$ & $0.68\pm 0.06$ & $0.08\pm 0.06$ & $0.74\pm 0.06$  \\
    \bottomrule
  \end{tabular}
\end{wraptable}

Results in Table \ref{tab:results} show moderate accuracy (RMSE$<$0.9 and r$^2\in[0.55,0.70]$) and low bias (ME$\leq 0.15$) levels for all developed models of IDP, either regionally, disaster-specific or globally. This is a necessary observation before the more ambitious goal of model interpretability. 
In this section we analyze the $\gamma_d$ rankings in a context of displacement at the different scales. 

\subsection{Climate-induced migrations in Africa and Asia}

\begin{wrapfigure}{r}{5cm}
\vspace{-0.5cm}
\centering
  \includegraphics[width=5cm]{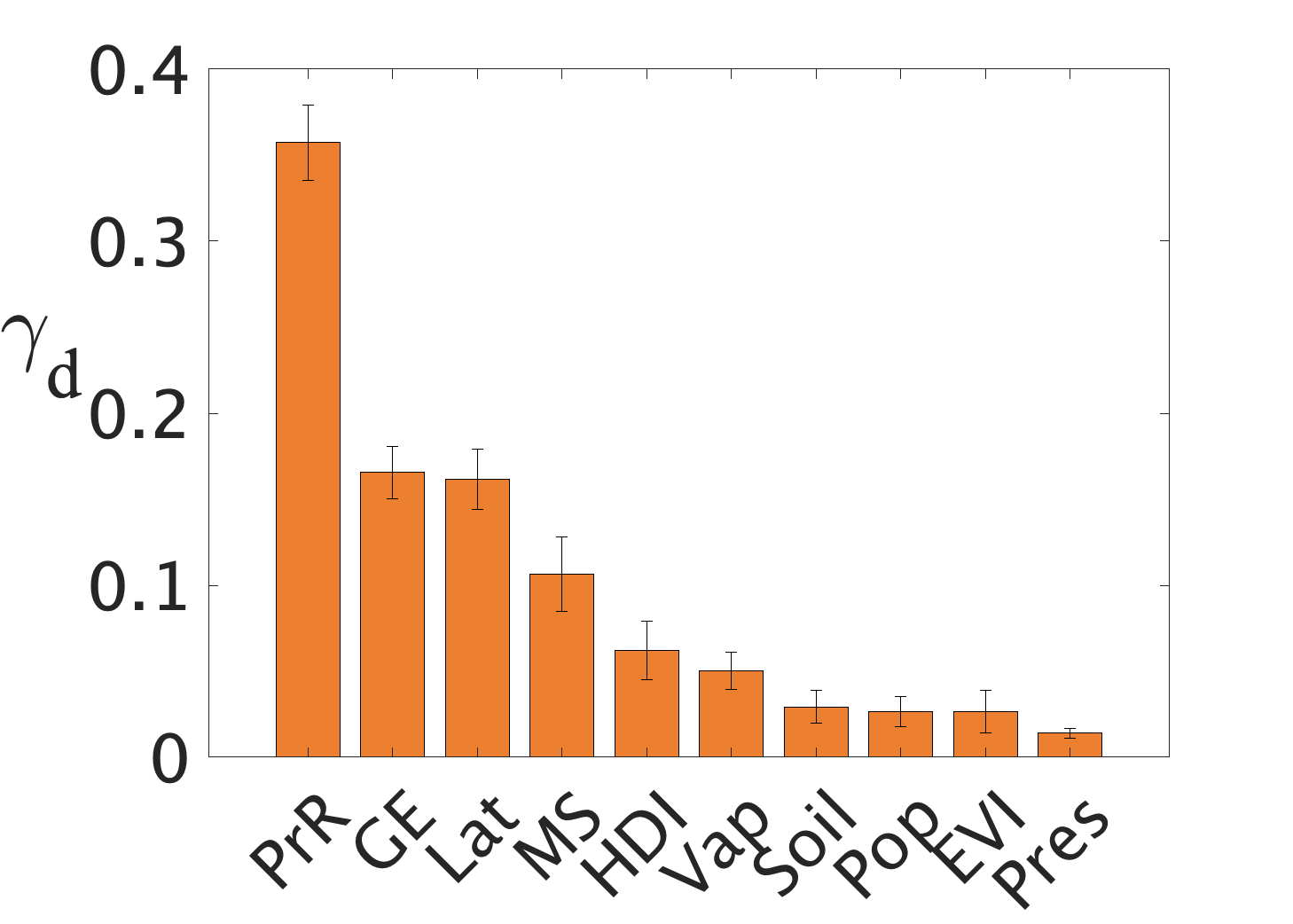}
\includegraphics[width=5cm]{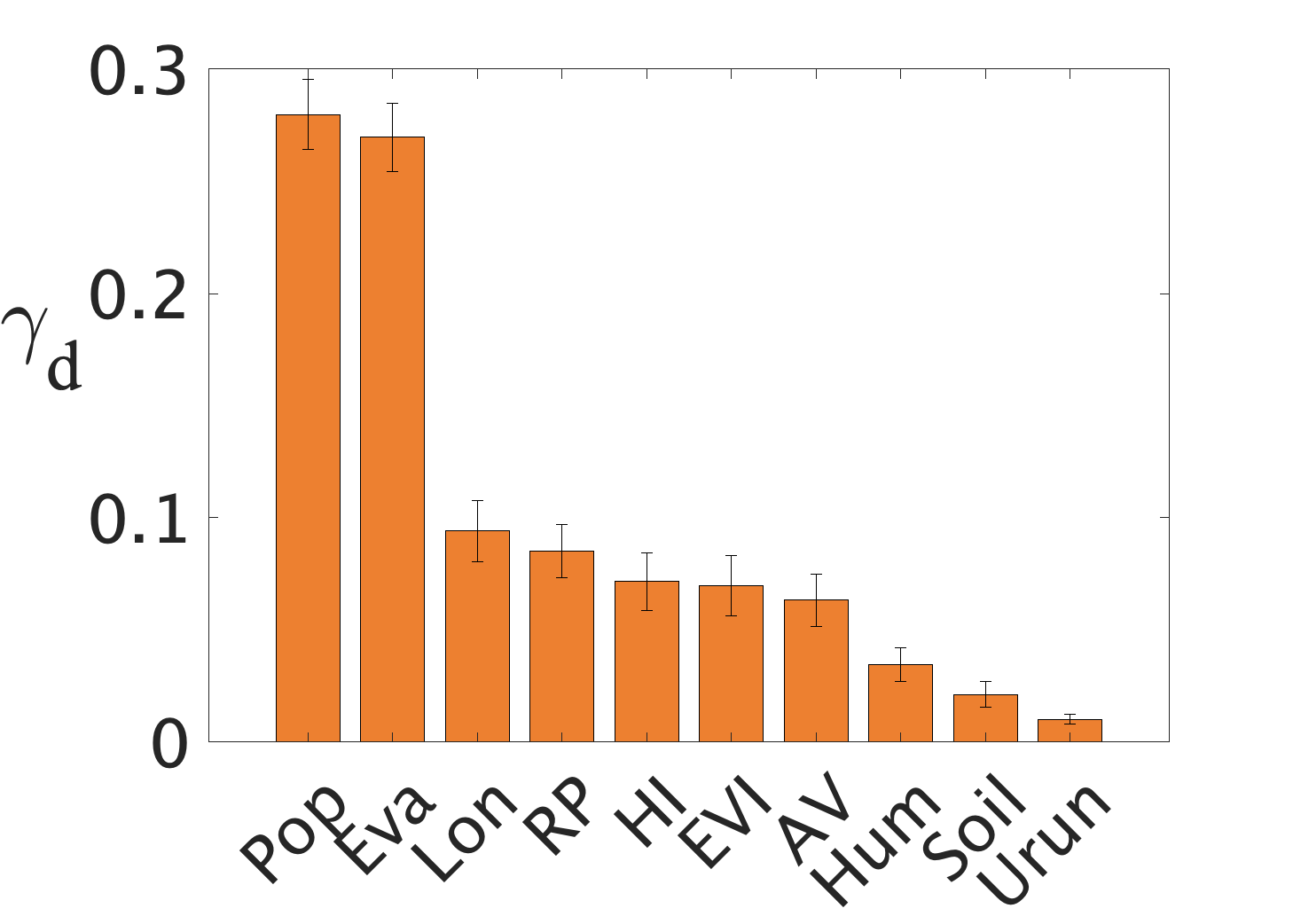}
  \caption{Ranking of climatic covariates inducing migration in Africa (top) and Asia (bottom) .\label{fig:regional}}
\end{wrapfigure}
We study here Africa and Asia as the most important regions of forced migration. Two individual models were thus developed for each continent. The ranking of covariates extracted by the GP model for migrations in Africa is shown in Fig.~\ref{fig:regional}[top] and indicates that precipitation rate ({\sf \small PrR}) is the most relevant factor, answering to the great impact of rain across the African continent. Both government effectiveness ({\sf \small GE}) and mortality due to lack of sanitation ({\sf \small MS}) follow, which are related to the socio-economic vulnerability of countries such as South Sudan, Sudan, DR Congo, Somalia or Nigeria and that also present a significant number of IDPs.  Another relevant factor is latitude ({\sf \small Lat}), which could account for the high displacement magnitudes observed above the Equator. An interesting finding is that {\sf \small Pop} scores little importance, suggesting that displacement magnitude in Africa does not relate to the most populated areas but to those which climate disasters have been more devastating.

Results for Asia in Fig.~\ref{fig:regional}[bottom] suggest that {\sf \small Pop} is the most relevant feature to explain the IDP in the region. This result is supported by the fact that 84\% of total IDPs in Asia come from $78$ disaster events in Bangladesh, China, India and Philippines alone, which represent tremendous population densities. Among climate features, evapotranspiration anomalies ({\sf \small Eva}) is by far the most relevant one. We hypothesize here that this is related to the  evaporation of water accumulations after heavy flood and storm extreme events, which could be possibly due to the impact of Monsoon floods in the most populated cities. Monitoring population density and evapotranspiration are suggested to effectively quantify IDP in Asia.

\subsection{Climate-induced migrations by flood and storm disasters}

\begin{wrapfigure}{r}{5cm}
\vspace{-0.5cm}
\centering
\includegraphics[width=5cm]{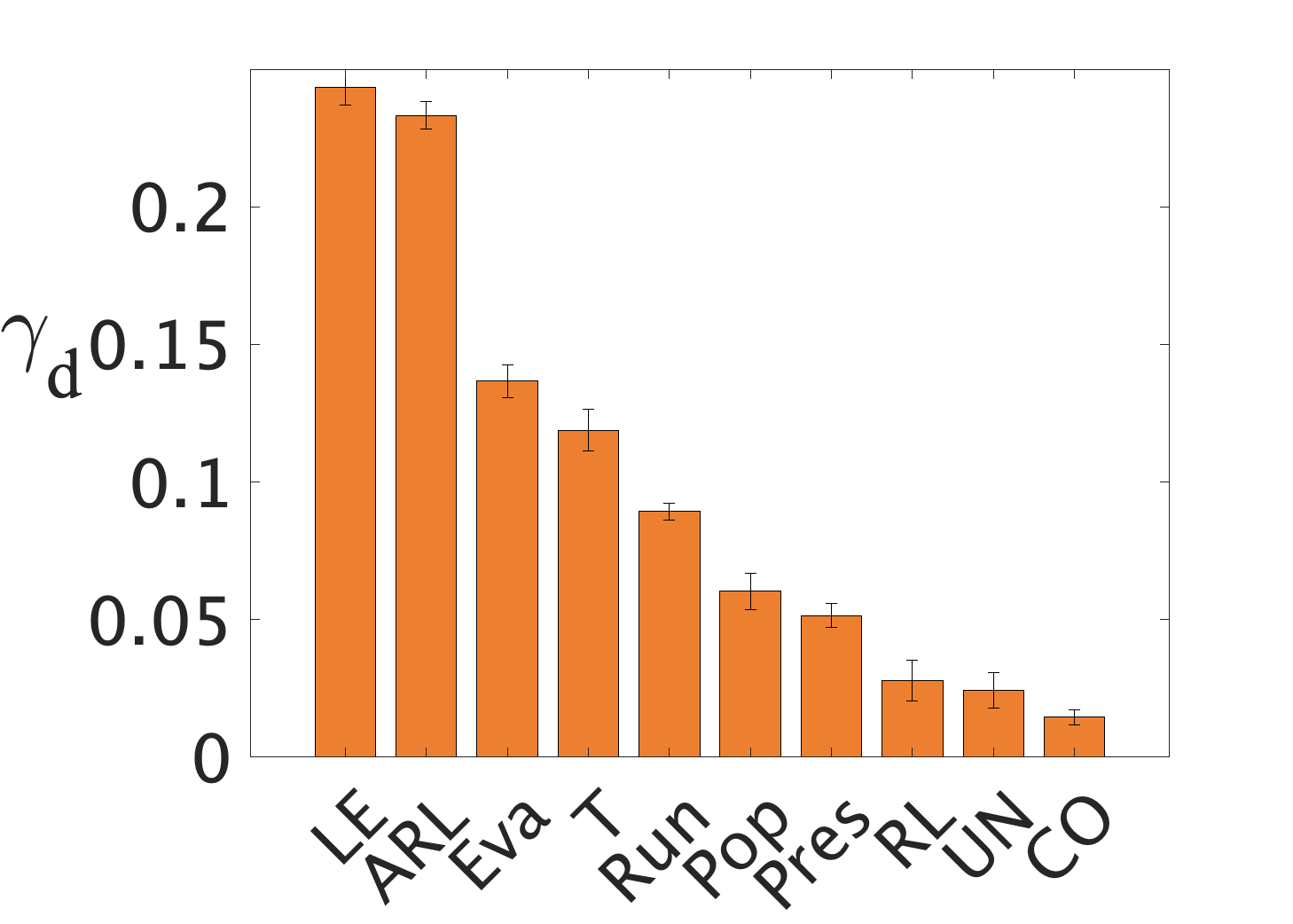} 
\includegraphics[width=5cm]{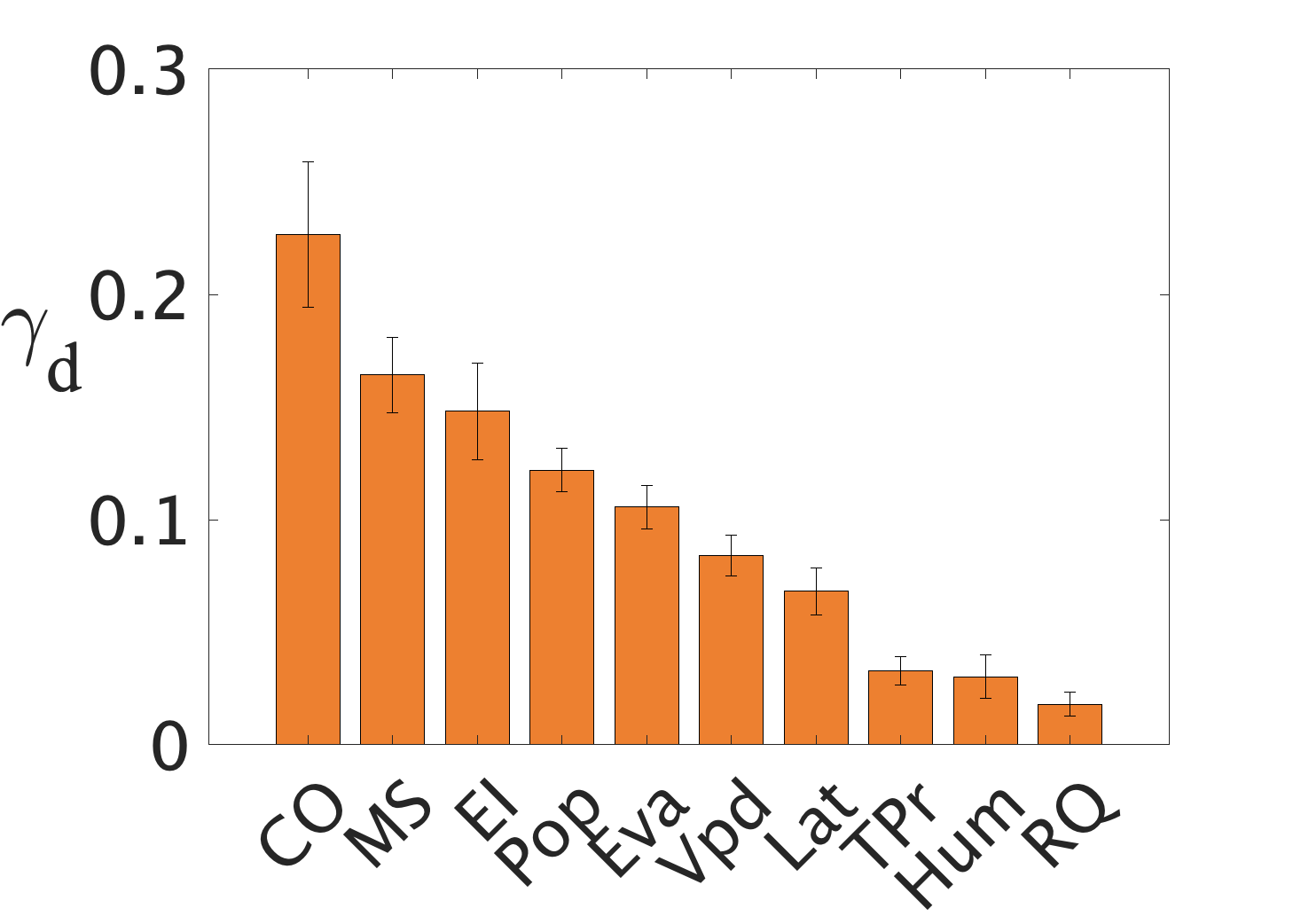} 
  \caption{Ranking of features obtained for flood and storm induced migrations. \label{fig:disasters}}
    \vspace{-12pt}  
\end{wrapfigure}
The previous regional studies could be affected by the type of disaster. Let us now study the factors impacting migrations induced by floods and storms separately. Flood modelling in Fig.~\ref{fig:disasters}[top] reveals that life expectancy ({\sf \small LE}) and percentage of arable land ({\sf \small ARL}) are the most dominant factors driving migrations after floods, and indicate that the GP model learned to associate IDP magnitude with specific countries. Maximum temperature ({\sf \small T}) importance could also be interpreted in this regard, taking into account temperature differences between Asia and Africa regions. For this reason, {\sf \small LE} importance must not be interpreted as a socio-economic vulnerability driving flood displacement. Nonetheless, the most severe IDP flood events occur in countries such as China, Philippines, Bangladesh, India, Nigeria or Ethiopia which have high {\sf \small ARL} values. This result could be displaying the effect of intensive agricultural practices, land degradation and deforestation on flood risk as a global scale problem\cite{ipccland, VEG_FLOOD}. 

Storm events in Bangladesh, China, India and Philippines account for 80\% of the IDPs caused by storms in our datataset. 
The storm model, cf. Fig.~\ref{fig:disasters}[bottom], must be interpreted on these terms. Results present the highest relevance for length of coastline in km ({\sf \small CO}) feature accounting for the severe impact of tropical cyclones in the coastal regions of China, India, Bangladesh and Philippines, but also the severe cyclones in Mozambique (Idai cyclone) and Vietnam (e.g. Vinta-Tembin cyclone), before dissipating inland. {\sf \small Pop} and {\sf \small Lat} features arise again, mainly due to the high IDP magnitudes present in South Asia, while {\sf \small Eva} and Water Vapour Deficit ({\sf \small VPD}) are variables associated to floods and cyclone forming conditions, respectively.

\subsection{Climate-induced migrations globally}

\begin{wrapfigure}{r}{5cm}
\vspace{-1cm}
\centering
\includegraphics[width=5cm]{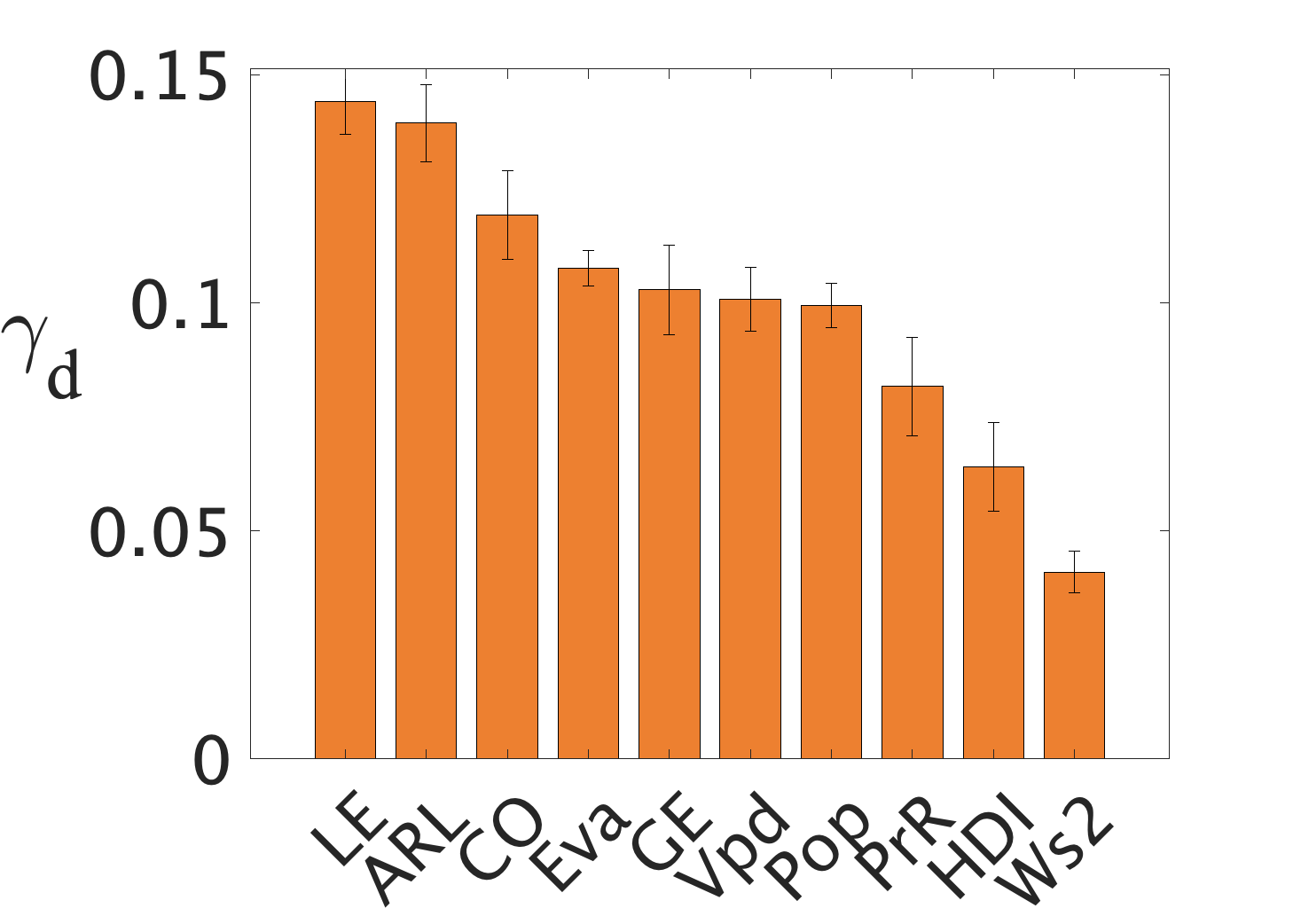}
  \caption{Ranking of features obtained for climate-induced global migrations.\label{fig:global}}
    \vspace{-12pt}  %hack to lift text up
\end{wrapfigure}
Global model results, cf. Fig.~\ref{fig:global}, are consistent with flood and storm models returning highest feature relevance for {\sf \small LE}, {\sf \small ARL} and {\sf \small CO} features which account for the most affected regions by flood and storms. Next, {\sf \small Eva} and {\sf \small Vpd} appear as the most relevant climate predictors of global displacements accounting for flood and storm events alongside with {\sf \small GE}, {\sf \small Pop} which appear in Asia and Africa Models. As many events and regions are combined together in the same model, the relative relevance of covariates is more even, yet still reflecting on the great importance of the combination of societal (life expectancy) and economical (land arable) factors.

\section{Conclusions}
 
We introduced the use of Gaussian processes in the challenging problem of interpreting the relative relevance of climate-induced human migrations. We collected and harmonized a database of meteorological, satellite-derived and socio-economic indicators to undertake our investigations. Results at regional, global and disaster-specific (floods and storms) showed the high importance of structural (socio-economic) covariates  as the driving forces to determine the magnitude of displacements. The results obtained have both societal and economic implications in disaster management and monitoring, which should be not only type but also country-specific
 
Despite these encouraging results there are a number of limitations to sort out in the future. Catastrophes are not the only drivers triggering forced migrations. It is widely accepted that turning points and complex saturation effect exist on life conditions \cite{inflexion}, which should be considered too. Actually, accumulation of previous disasters can be a discriminative factor, but inclusion requires longer time series. This is a challenging problem with a high level of noise and uncertainty, and with a great number of involved counfounders, where causal discovery  could -and should- say a word.

\acksection

Research partly funded by the `Ayudas Fundaci\'on BBVA a equipos de investigaci\'on cient\'ifica 2019' in the project {\em `Causal inference in the human-biosphere coupled system',}  and the EU H2020 DeepCube {\em `Explainable AI pipelines for big Copernicus data'}.
Gustau Camps-Valls were supported by the European Research Council (ERC) under the ERC-CoG-2014 SEDAL (647423) project.

%\bibliography{references}
%\bibliographystyle{unsrt}

\clearpage
\section*{Appendix I. Details on the IDP database}
\label{app:events}

\begin{table}[H]
  \caption{IDP events by country, disaster and regions}
  \label{sample-table}
    \centering
  \begin{tabular}{llllll}
    \toprule
    \multicolumn{2}{c}{} &   \multicolumn{2}{c}{Flood} &
    \multicolumn{2}{c}{Storm} \\
    \cmidrule(r){3-4}
    \cmidrule(r){5-6}
    Country     & Region     & Events & IDPs & Events & IDPs \\
    \toprule
    Afghanistan & Asia  & $13$ & $169\,700$ & $0$ & $0$   \\
    
    Bangladesh & Asia  & $8$ & $791\,600$ & $4$ & $4\,263\,000$   \\
    
    Cambodia & Asia  & $3$ & $48\,300$ & $2$ & $57\,300$   \\
    
    China & Asia  & $11$ & $4\,834\,000$ & $15$ & $6\,130\,000$   \\
    
    India & Asia  & $7$ & $5\,987\,500$ & $7$ & $2\,800\,600$   \\
    
    Iraq & Asia  & $3$ & $81\,100$ & $0$ & $0$   \\
   
    Laos & Asia  & $0$ & $0$ & $2$ & $120\,000$   \\
    
    Myanmar & Asia  & $3$ & $829\,000$ & $6$ & $42\,600$   \\
    
    Nepal & Asia  & $3$ & $483\,900$ & $2$ & $15\,700$   \\
    
    Philipines & Asia  & $14$ & $2\,090\,300$ & $10$ & $4\,174\,800$   \\
    
    Thailand & Asia  & $3$ & $45\,700$ & $2$ & $60\,000$   \\
   
    Sri-Lanka & Asia  & $6$ & $284\,400$ & $0$ & $0$   \\
    
     Vietnam & Asia  & $2$ & $26\,000$ & $8$ & $780\,000$   \\
   
    Angola & Africa  & $8$ & $23\,700$ & $1$ & $1\,600$   \\
  
    Congo & Africa  & $2$ & $166\,000$ & $0$ & $0$   \\
  
    DR Congo & Africa  & $11$ & $251\,400$ & $3$ & $40\,100$   \\
    
    Ethiopia & Africa  & $8$ & $587\,600$ & $0$ & $0$   \\
   
    Kenya & Africa  & $6$ & $438\,000$ & $0$ & $0$   \\
 
    Mali & Africa  & $2$ & $5\,700$ & $0$ & $0$   \\
  
    Mozambique & Africa  & $2$ & $25\,100$ & $4$ & $676\,000$   \\
   
    Niger & Africa  & $4$ & $348\,300$ & $7$ & $3\,999$   \\
  
    Nigeria & Africa  & $8$ & $842\,200$ & $3$ & $8\,100$   \\
  
    Rwanda & Africa  & $3$ & $58\,000$ & $2$ & $3\,800$   \\
  
    Somalia & Africa  & $4$ & $709\,100$ & $1$ & $13\,100$   \\
    
    South Sudan & Africa  & $4$ & $361\,000$ & $0$ & $0$   \\
    
    Sudan & Africa  & $4$ & $446\,800$ & $0$ & $0$   \\
   
    Uganda & Africa  & $7$ & $333\,000$ & $0$ & $0$   \\

    \bottomrule

{\bf Africa} &  & $79$ & $4\,708\,400$ & $14$ & $742\,700$ \\   
{\bf Asia} &  & $70$ & $15\,577\,200$ & $66$ & $21\,141\,000$ \\  
{\bf Total} &   & $149$ & $20\,285\,600$ & $80$ & $21\,883\,700$ \\   
    \bottomrule
  \end{tabular}
\end{table}

\clearpage
\section*{Appendix II. Details on the covariates}
\label{app:variables}
\begin{center}

\begin{longtable}{p{3cm}p{1cm}p{2cm}p{4cm}p{3cm}}
\toprule \toprule
{\bf Feature} & {\bf Short} & {\bf Years} & {\bf Description} & {\bf Reference} \\\midrule
\multicolumn{5}{l}{{\bf Domain: \textit{Migration}}}\\ 
\midrule
Internally Displaced Person & IDP & 2017-2019 & Person forced to leave home, in particular as a response of a climate disaster and that has not crossed any international border. & IDMC website \cite{IDMCweb} \\
\midrule
\multicolumn{5}{l}{\textbf{Domain: \textit{Atmospheric}}}\\ 
\midrule
Evapotranspiration & Eva & 2017-2019 & Evapotranspiration from VIC4.1.2 (and Noah3.3) is the sum of three weighted components by their surface contribution: canopy intercepted evaporation, vegetation transpiration and earth evaporation. Units: kg m-2 s-1. & FLDAS Noah Land Surface Model L4 Global Monthly 0.1 x 0.1 degree \cite{FDLAS1} \\
\midrule
Specific Humidity & Hum    & 2017-2019 & Units: kg kg-1 & FLDAS Noah Land Surface Model L4 Global Monthly 0.1 x 0.1 degree \cite{FDLAS1} \\
\midrule
Total Precipitation Rate            & PrR & 2017-2019 & Monthly total precipitation rate combining data from GDAS, MERRA-2 and CHIRPS. Units: kg m-2 s-1                                 & FLDAS Noah Land Surface Model L4 Global Monthly 0.1 x 0.1 degree \cite{FDLAS1}                      \\ \midrule
Wind Speed 1               & Ws1    & 2017-2019 & Wind speed at surface level (10m). Units: m s-1                                                                                                                           & FLDAS Noah Land Surface Model L4 Global Monthly 0.1 x 0.1 degree \cite{FDLAS1}   \\
\midrule
Near Surface Temperature (2m) & T      & 2017-2019 &  Units: K                                                                                   & FLDAS Noah Land Surface Model L4 Global Monthly 0.1 x 0.1 degree \cite{FDLAS1}.                                                                                            \\ \midrule

Water Vapour Pressure                    & Vap    & 2017-2019 & Measure of the vertical column of water vapour Measure combining meteorological station datas of  WorldClim v2, CRU Ts4.0 and JRA-55 Units: kPa                                                                                                                                                                                                                                                                   & TerraClimate: Monthly Climate and Climatic Water Balance for Global Terrestrial Surfaces 2.5 arc minutes \cite{terra} \\
\midrule
Water Vapour Deficit                 & Vpd & 2017-2019 & Measure of monthly water deficit around mean value of the dataset combining meteorological station datas of  WorldClim v2, CRU Ts4.0 and JRA-55 Units: kPa                                                            & TerraClimate: Monthly Climate and Climatic Water Balance for Global Terrestrial Surfaces 2.5 arc minutes \cite{terra}                                                                                                                                                                                              \\ \midrule

Precipitation Accumulation                & PrA    & 2017-2019 & Measure of monthly accumulated precipitation combining meteorological station datas of  WorldClim v2, CRU Ts4.0 and JRA-55 Units: mm                                                                                                                                                                                                                                 & TerraClimate: Monthly Climate and Climatic Water Balance for Global Terrestrial Surfaces 2.5 arc minutes \cite{terra}                                                                                                                                                                                          \\ \midrule

Mean Total Precipitation               & TPr  & 2017-2019 & Precipitation rate over earth surface if it was uniformly distributed over the selected region. Precipitation is generated following ECMWF "Integrated Forecasting System" (IFS) models. Units: kg m-2 s-1                                                                                                            & ERA5 Monthly aggregates 0.25 arc degrees -  ECMWF / Copernicus Climate Change Service \cite{era5}                                                                                                                                                                                                                 \\ \midrule

Wind Speed 2                       & Ws2    & 2017-2019 & Horizontal wind component at 10m. Unidades: m s-1                                                                                                                                                                                                                                    & ERA5 Monthly aggregates 0.25 arc degrees -  ECMWF / Copernicus Climate Change Service \cite{era5}                                                                                                                                                                                                              \\ \midrule

Sea Level Pressure                           & Pres    & 2017-2019 & Minima of atmospheric pressure adjusted to sea level altitude. Units: Pa.                                                                                                                                                & ERA5 Monthly aggregates 0.25 arc degrees -  ECMWF / Copernicus Climate Change Service \cite{era5}     
       \\ \midrule                                                       
\multicolumn{5}{l}{\textbf{Domain: \textit{Earth}}}\\ 
\midrule
       
Soil Moisture                         & Soil   & 2017-2019 & Soil moisture at 0 - 10 cm underground. Units: m\textasciicircum{}3 m-3                                                                                                                                                                        & FLDAS Noah Land Surface Model L4 Global Monthly 0.1 x 0.1 degree \cite{FDLAS1}                                                                                                                                                                                                             \\ \midrule
Water Run-off                 & Run    & 2017-2019 & In Noah3.3, water Run-off is computed using the two layers conceptual scheeme of Schaake et al.37 based on a simple water balance model. Units: kg m-2 s-1 & FLDAS Noah Land Surface Model L4 Global Monthly 0.1 x 0.1 degree \cite{FDLAS1}                                                                                                                                                                                                             \\ \midrule
Underground Water Run-off          & Urun   & 2017-2019 & Units: kg m-2 s-1                                                                                                                                                                                                                                                                                                                                                                                                                                                                   & FLDAS Noah Land Surface Model L4 Global Monthly 0.1 x 0.1 degree \cite{FDLAS1}                                                                                                                                                                                                             \\ \midrule
\multicolumn{5}{l}{{\textbf{Domain: \textit{Demographic}}}}\\ 
\midrule
Total Population                & Pop    & 2015    & Population density estimate based on national registers adjusted by the "United Nation's World Population Prospects" (UN WPP) census in 2015 multiplied by the area of the affected region  & GPWv411: Population Density (Gridded Population of the World Version 4.11) \cite{pop} \\ \midrule

Rural Population                     & RP     & 2018         & percentage of Rural Population the country.                                                                                                                                                                                                                                                                                                                                                                               &     United Nations Development Programme: Human Development Reports \cite{UN}                                                                                                                                                                                                                                                                                                      \\ \midrule
Birth Rate                  & BR     & 2018         &   Birth Rate in the country.                                                                                                                                                          &United Nations Development Programme: Human Development Reports \cite{UN}                                                                                                                                                                                                                                                                                                           \\ \midrule
Life Expectancy                      & LE     & 2018         &    Average Life Expectancy in the country.                                                                                                                                                                                                                                                                                         & United Nations Development Programme: Human Development Reports \cite{UN}                                                                                                                                                                                                                                                                                               \\ \midrule
Median Age                           & MA     & 2018         & Median Age of the population of the country &  United Nations Development Programme: Human Development Reports \cite{UN}  
\\ \midrule                                                                                                                                           Unemployment                                 & UN     & 2018         & Unemployment percentage in the country.                                                                                                                                                                                                                                                                                   &  United Nations Development Programme: Human Development Reports \cite{UN}                                                                                                                                                                                                                                                                                                         \\ \midrule
\multicolumn{5}{l}{{\textbf{Domain: \textit{Socio-economic variable}}}}\\ 
\midrule
Mortality due to lack of sanitation          & MS     & 2018         & Number of deaths due to poor hygiene and sanitary services per 100.000 inhabitants in the country.                                                                                                                                                                                                                                                                                                                                                                                                                                                             &           United Nations Development Programme: Human Development Reports \cite{UN}                                                                                                                                                                                                                                                                                                \\ \midrule
GDP per Capita                        & GDP    & 2018         & FMI establishes that ``GDP measures the monatary value of final goods and services, that are bought by the final user, produced in a country per capita''                                                                                                                                                                                                                                                                                               &    World Development Indicator \cite{WDI}                                                                                                                                                                                                                                                                                                        \\ \midrule
Exports \& Imports                    & EI     & 2018         &    It's a measurement on the trade of a country. measures the total exports and imports in dollars as a \% of the GDP of the country.                                                                                                                                                                                               &       United Nations Development Programme: Human Development Reports \cite{UN}                                                                                                                                                                                                                                                                                                    \\ \midrule
Electricity Access                & AE     & 2018         &   Percentage of the population with access to electricity                                                                                                                                                                                                      &    United Nations Development Programme: Human Development Reports \cite{UN}                                                                                                                                                                                                                                                                                                       \\ \midrule
Cereal Yield                        & CY     & 2018         &    Kg of cereals produced in the country.                                                                                                                                                                                                          &                 United Nations Development Programme: Human Development Reports \cite{UN}                                                                                                                                                                                                                                                                                          \\ \midrule
Employment in Agriculture                               & EA     & 2018         &  Fracction of total employment that belong to agricultural practises.                                                                                                                                                                                                                                                                                                 &                       United Nations Development Programme: Human Development Reports \cite{UN}                                                                                                                                                                                                                                                                                    \\ \midrule
Paved Road                            & PR     & 2018         & km of paved road in the country.                                                                                                                                                                                                                     &      United Nations Development Programme: Human Development Reports \cite{UN}                                                                                                                 \\ \midrule
\multicolumn{5}{l}{{\textbf{Domain: \textit{Socio-economic Index}}}}\\ 
\midrule
Human Inequality                              & HI     & 2018         &     Measures the capability of citizens of accessing the same life conditions in a country if inequality conditions did not exist. &  World Development Indicator \cite{WB}                                                                                                                                                      \\ \midrule
Concentration Index                              & CI     & 2018         &  Defined as twice the area below the concentration curve, $L(p)$, and the line of equality. In case of not existing salary equality CI is equal to 0.                                                                                                                                                  & World Development Indicator \cite{WB}                                                                                                                                                                                          \\ \midrule
Education Index                       & EDI    & 2018         &  50\% weighted average between the mean years of adult scholarization and the mean years of children scholarization.                                                                                                                                                                                                                                                                       & World Development Indicator \cite{WB}                                                                                                                                                                                                \\ \midrule
Human Development Index               & HDI    & 2018         &  Summarizes a measurement of the development in the country: A health, life expectancy, education and life quality.HDI is the geometric average between life expectancy, Education Index and Income Index.                                                                                                                                                                               &  World Development Indicator \cite{WB}                                                                                                                   \\ \midrule
Income Index                          & II     & 2018         & Income Index o GNI per capita is defined as the GDP plus foreign money influx minus the money leaving the country per capita.                                                                &         World Development Indicator \cite{WB}                                                                                                                                                                                             \\ \midrule
Human Capital Index                   & HCI    & 2018         &            The index measures the capability of the country to mobilize the economic and profesional potential of its citizens. It also measures the capital loses trough a lack of education and sanitation.                                                                                                                                                                   &      World Development Indicator \cite{WB}                                                                                                        \\ \midrule
\multicolumn{5}{l}{\textbf{Domain: \textit{Environmental}}}\\ 
\midrule
Coastline                            & CO     & Static       &     Length of coastline in the area in km.                                                                                                                                                                                                                         &                   United Nations Development Programme: Human Development Reports \cite{UN}                                                                                                                                                                                  \\ \midrule
Latitude                              & LAT    & Static     & Latitude &    United Nations Development Programme: Human Development Reports \cite{UN}                                                                                                                                                                                                                                                                                                                                                  \\ \midrule
Longitude                           & LON    & Static  &   Longitude  &   United Nations Development Programme: Human Development Reports \cite{UN}                                                                                                                                                                                                                                                                                                                                                                                                                                                                         \\ \midrule
Altitude                            & AL     & 2018         & Average altitude in the affected area. Units: m                                                    & SRTM Digital Elevation Data Version 4. 90m Resolution.  http://srtm.csi.cgiar.org/     \\ \midrule                                                                                                                                                                   
Arable Land & ARL    & 2018         & Percentage of land cultivated for crops like wheat, maize, and rice that are replanted after each harvest in a country.                                                                                                                                                                  &  United Nations Development Programme: Human Development Reports \cite{UN}                                                                                                                                                                                                                                                                                                         \\ \midrule
Enhanced Vegetation Index & EVI  & 2017-2019 & EVI is an enhanced vegetation index designed to improve vegetation signal in regions with high biomass by reducing atmospheric effects. Used criteria for the value is low clouds, low vision angle and the highest value of EVI. & MOD13A2.006 Terra Vegetation Indices 16-Day Global 1km.
\cite{modis} \\
\midrule
\multicolumn{5}{l}{{\textbf{Domain: \textit{Political Index}}}}\\ 
\midrule
Corruption Index                      & COR    & 2018         &        Captures the degree of perception of which public charges use public capital for private interests                                                                                                                                                                                          &                      Worldwide Governance Indicators \cite{WFI}                                                                                                                                                                                                                                                                                   \\ \midrule
Goverment Effectiveness                        & GE     & 2018         &    Captures the perceptions on the quality of public services, civil services and their independence from political pressures as well as the perception on credibility, compromise and quality of govermental policies.                                                                                                                                                                                                                                                                                                                                                                                 &          Worldwide Governance Indicators \cite{WFI}

\\ \midrule
Absence of Violence                                   & AV     & 2018         &   Measures the perception and cases on political instability and absence of violence and terrorism.                                                                                                                                                                                               &          Worldwide Governance Indicators \cite{WFI}                                                                                                                                                                                                                                                                                               \\ \midrule
Regulatory Quality                    & RQ     & 2018         &      Captures the perception of the ability of a goverment to enforce effective policies and regulations.                                                                                              &           Worldwide Governance Indicators \cite{WFI}                                                                                                                                                                              \\ \midrule
Rule of Law                           & RL     & 2018         & Captures perception on the quality of laws and the degree of fairness in which the law is applied   &                Worldwide Governance Indicators \cite{WFI}                                                                                                                 \\ \midrule
Voice Accountability                  & VA     & 2018         &   Captures the perceptions on the level in which citizens from a country can actively engage with the selection of its goverment, freedom of speech, association and media.  &                                   Worldwide Governance Indicators \cite{WFI}                                                                                                                                                   \\ \bottomrule \bottomrule                               
\end{longtable}

\end{center}

\end{document}